%% file: main.tex
\documentclass[runningheads]{llncs}

\input{flow/_flow}

\begin{document}

\input{flow/_commands}

\input{head/title}

\input{head/authors}

\maketitle 
\input{head/abstract}
\input{body/_body}

%
%
%
\bibliographystyle{splncs04}

\bibliography{mybib}

\clearpage

\input{body/supplement.tex}

\end{document}

%% file: flow/_flow.tex
\newcommand*\samethanks[1][\value{footnote}]{\footnotemark[#1]}

\usepackage[symbol]{footmisc}

\usepackage[table,xcdraw]{xcolor}
\usepackage{float}
\usepackage{caption}
\usepackage{subcaption}
\usepackage{graphicx}
\usepackage[T1]{fontenc}
\usepackage{numprint}
\usepackage{newclude}
\usepackage{siunitx}
\usepackage{amssymb}
\usepackage{tabularx}
\usepackage{add_pack/tabularray}
\usepackage{add_pack/ninecolors}
\usepackage{hyperref}
\hypersetup{colorlinks,allcolors=black}
\usepackage{navigator}
\usepackage{amsmath}
\usepackage{rotating}
\usepackage{multirow}
\usepackage{cleveref}
\usepackage[nolist]{acronym}
\input{flow/acronyms.tex}

\makeatletter
\newcommand*{\centerfloat}{%
  \parindent \z@
  \leftskip \z@ \@plus 1fil \@minus \textwidth
  \rightskip\leftskip
  \parfillskip \z@skip}
\makeatother

%% file: flow/acronyms.tex
\begin{acronym}
\acro{PP}{Pre-Post-OP}
\acro{PI}{Pre-Intra-OP}
\acro{VM}{VoxelMorph}
\acro{MR}{Magnetic Resonance}
\acro{MRI}{Magnetic Resonance Imaging}
\acro{ANTs}{Advanced Normalization Tools}
\acro{BraTS-Reg}{Brain Tumor Sequence Registration Challenge}
\acro{DRN}{Deformable Registration Network}
\acro{IO}{Instance-Specific Optimization}
\acro{MSE}{Mean Squared Error}
\acro{TRE}{Target Registration Error}
\acro{NCC}{Normalized Cross Correlation}
\acro{CNN}{Convolutional Neural Network}
\acro{DL}{Deep Learning}
\acro{ML}{Machine Learning}
\acro{SyN}{Symmetric Normalization}
\acro{OP}{operative procedure}
\acro{FFD}{Free-Form Deformation}
\acro{T1}{native T1-weighted}
\acro{T1CE}{contrast-enhanced T1}
\acro{T2}{T2-weighted}
\acro{FLAIR}{T2 Fluid Attenuated Inversion Recovery}
\end{acronym}

%% file: flow/_commands.tex
\newcommand{\eac}[1]{\emph{\ac{#1}}}
\newcommand{\eacp}[1]{\emph{\acp{#1}}}
\newcommand{\eacf}[1]{\emph{\acf{#1}}}

%% file: head/title.tex
\title{Primitive Simultaneous Optimization\\ of Similarity Metrics for Image Registration}

\titlerunning{Primitive Simultaneous Optimization of Similarity Metrics}

%% file: head/authors.tex
\author{Diana Waldmannstetter\inst{1,2}\thanks{diana.waldmannstetter@tum.de} 
\and Benedikt Wiestler\inst{3} 
\and Julian Schwarting\inst{3} 
\and Ivan Ezhov\inst{1,4} 
\and Marie Metz\inst{3}
\and Spyridon Bakas\inst{5,6,7}
\and Bhakti Baheti\inst{5,6,7}
\and Satrajit Chakrabarty\inst{8}
\and Daniel Rueckert\inst{9,10}  
\and Jan S. Kirschke\inst{3} 
\and Rolf A. Heckemann\inst{11}
\and Marie Piraud\inst{12} 
\and Bjoern H. Menze\inst{2}\thanks{equal contribution}
\and Florian Kofler\inst{1,3,4,12}\samethanks}

\authorrunning{D. Waldmannstetter et al.}

\institute{Department of Informatics, Technical University of Munich, Munich, Germany 
\and Department of Quantitative Biomedicine, University of Zurich, Zurich, Switzerland 
\and Department of Diagnostic and Interventional Neuroradiology, School of Medicine, Klinikum rechts der Isar, Technical University of Munich, Munich, Germany 
\and TranslaTUM - Central Institute for Translational Cancer Research, Technical University of Munich, Munich, Germany
\and Center for Artificial Intelligence and Data Science for Integrated Diagnostics (AI2D) and Center for Biomedical Image Computing and Analytics (CBICA), University of Pennsylvania, Philadelphia, PA, USA
\and Department of Pathology and Laboratory Medicine, Perelman School of Medicine, University of Pennsylvania, Philadelphia, PA, USA
\and Department of Radiology, Perelman School of Medicine, University of Pennsylvania, Philadelphia, PA, USA
\and Department of Electrical and Systems Engineering, Washington University in St.Louis, St. Louis, MO, USA
\and Artificial Intelligence in Healthcare and Medicine, Technical University of Munich, Munich, Germany
\and Department of Computing, Imperial College London, London, U.K.
\and Department of Medical Radiation Sciences, University of Gothenburg, Gothenburg, Sweden
\and Helmholtz AI, Helmholtz Zentrum München, Munich, Germany}

%% file: head/abstract.tex
\begin{abstract}
Even though simultaneous optimization of similarity metrics is a standard procedure in the field of semantic segmentation, surprisingly, this is much less established for image registration.
To help closing this gap in the literature, we investigate in a  complex multi-modal 3D setting whether simultaneous optimization of registration metrics, here implemented by means of primitive summation, can benefit image registration.
We evaluate two challenging datasets containing collections of pre- to post-operative and pre- to intra-operative \ac{MR} images of glioma.
Employing the proposed optimization, we demonstrate improved registration accuracy in terms of \eac{TRE} on expert neuroradiologists' landmark annotations.
\input{head/keywords}
\end{abstract}

%% file: head/keywords.tex
\keywords{Registration \and Brain Tumor \and Similarity Metric \and Loss Function \and Glioma}

%% file: body/_body.tex
\include*{body/01_introduction}

\include*{body/02_methods}

\include*{body/03_experiments_and_results}

\include*{body/04_discussion}

\include*{body/05_conclusion}

\include*{body/acknowledgement}

%% file: body/01_introduction.tex
\section{Introduction}
The standard treatment for glioma is an \eac{OP} aiming to remove the tumor in full.
Clinicians measure the success of surgery by comparing pre- and post-operative scans.
Furthermore, this comparison is required for subsequent treatment planning, such as radiation therapy.
Image registration techniques can enhance this process by providing a direct overlay of the differing structures.
Enormous tissue shift and consequential missing correspondences mark a common side effect of tumor resection and pose a major challenge in registering pre- to post-operative images.
The same holds true for intra-operative imaging in order to keep track of the surgery progress. 
A fast and accurate image registration method is beneficial for the precise estimation of tumor resection.
Additionally, image registration is an important part of the preprocessing for segmentation algorithms \cite{kofler2023we,kofler2023blob}. Moreover, registration can be part of the segmentation algorithm itself \cite{fidon2022dempster}.
Recent advances in the field of \eac{DL} also benefited medical image registration. 
Methods like \eac{VM} \cite{balakrishnan2019voxelmorph}, \emph{LapIRN} \cite{mok2020large}, and \emph{TransMorph} \cite{chen2022transmorph} have demonstrated that unsupervised deformable image registration is a promising alternative to frameworks based on iterative optimization algorithms like \eac{SyN} from \eac{ANTs} \cite{avants2009advanced}, \eac{FFD} \cite{rueckert1999nonrigid} or toolboxes like Elastix \cite{klein2009elastix}.
\eac{DL} methods provide comparable performance while significantly improving processing time. 
Within the scope of the \eac{BraTS-Reg} \cite{baheti2021brain}, there has also been considerable development of registration algorithms for \ac{MR} brain images before and after tumor resection \cite{canalini2022iterative,grossbrohmer2022employing,meng2022brain,mok2022robust,wodzinski2022unsupervised}. 
Unsupervised \eac{DRN} can nicely register healthy brain scans \cite{balakrishnan2019voxelmorph,chen2022transmorph,mok2020large}; however, they frequently struggle with large pathologies.
Therefore, \eac{IO} proved to be advantageous to mitigate major deformations \cite{mok2022robust,wodzinski2022unsupervised}.
Existing deep learning registration algorithms usually make use of a single similarity metric, often coupled with a smoothing regularization based on the deformation field, to be optimized in the loss function \cite{balakrishnan2019voxelmorph, mok2020large}.
However, there is only limited literature investigating the combination of multiple similarity metrics to improve registration results \cite{avants2008multivariate,avants2011reproducible,boukellouz2017evaluation,ferrante2017deformable,uss2020selection,wachs2009multi,zhou2015combined}.

In this work, we extend an unsupervised \eac{DRN} using a combination of image similarity metrics, which are optimized simultaneously in the loss function. 
We benchmark the performance of our approach against the baseline on two challenging datasets of pre- to post-operative as well as pre- to intra-operative brain tumor images. 
Furthermore, we compare our approach to established reference methods, achieving competitive results. 
Following \cite{balakrishnan2019voxelmorph,mok2022robust,wodzinski2022unsupervised}, we opt for instance-specific optimization in addition to the \eac{DRN}.
We demonstrate that the simultaneous optimization of two similarity metrics can improve registration accuracy in terms of \eac{TRE} on expert landmark annotations. 

%% file: body/02_methods.tex
\newpage
\section{Methods}
Our workflow consists of three steps.
First, we train a \eacf{DRN}, followed by \eacf{IO} at test time. 
Both \eac{DRN} and \eac{IO} optimize the same loss function. 
We then evaluate the registration performance in terms of \eacf{TRE}.

\subsection{Unsupervised Deformable Registration}
\begin{figure}[tbp]
\centering
\includegraphics[width=\textwidth]{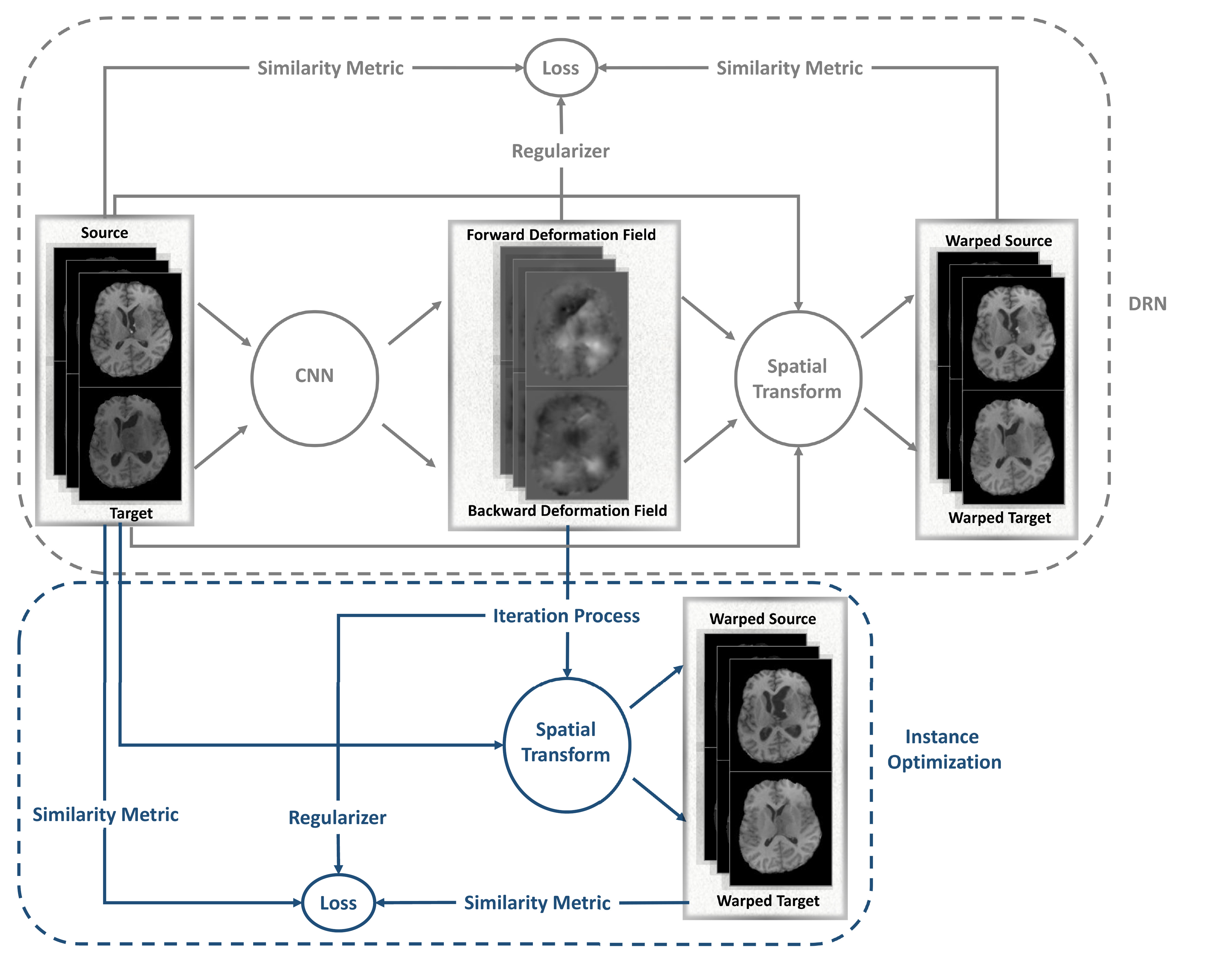}
\caption{Overview of the workflow. An unsupervised deformable registration network (\eac{DRN}, top) is combined with \eacf{IO} (bottom) for iterative refinement at test time using the output deformation field of the trained \eac{DRN}.
\eac{DRN} as well as \eac{IO} are trained bidirectionally, providing both forward and backward deformation fields.
For both modules, the loss function combines either a single or multiple image similarity metrics with a smoothness regularization on the deformation field.}
\label{fig:network}
\end{figure}

We start our approach with a \eac{DRN}, similar to \cite{balakrishnan2019voxelmorph}.
Given two 3D images, a source image $X$ and a target image $Y$, the network models the following function $f_\theta(X,Y)=u$, where the displacement field $u$, aligning $X$ and $Y$, is defined bidirectionally, leading to $u_{x,y}=f_\theta(X,Y)$ and $u_{y,x}=f_\theta(Y,X)$ with $\theta$ being a set of learning parameters.
$X$ and $Y$ are then warped with the respective displacement field using a spatial transform.
While many \eac{CNN}-based models are applicable here, like \cite{balakrishnan2019voxelmorph}, we opt for a \emph{U-Net}-like architecture with encoder, decoder and skip connections. 
The network architecture is based on an early implementation of \eac{VM}, see \href{https://github.com/voxelmorph/voxelmorph/blob/183fdf2a0426c6be187a0f6fbe24a7e05e026922/voxelmorph/torch/networks.py}{GitHub implementation}. 
The loss function comprises two main components, which are an image similarity metric and a smoothness regularizer for the displacement field. 
In addition to optimizing a single similarity metric as part of the loss function, we opt for optimizing multiple metrics simultaneously.
An overview of the workflow is shown in \Cref{fig:network}.

\subsection{Proposed Combined Loss Function}
We define the objective function $L$ for bidirectional training as follows:
\begin{equation}
    L=\sum_{n=1}^{N} \left( L_{Sim_{n}}^{forward}\cdot\omega_{n}+L_{Sim_{n}}^{backward}\cdot\omega_{n}\right )+L_{Reg}\cdot\lambda
\end{equation}
where $N$ is the number of similarity metrics $L_{Sim}$,  $L_{Reg}$ is the regularization term, and the respective weights are denoted by $\omega_{n}$ and $\lambda$.

\subsection{Instance-specific Optimization}
Since the registration results of a plain \eac{DRN} might not always be sufficient, especially in the case of pathologies, \eac{IO} is added in order to further refine the registration. 
Similar to \cite{balakrishnan2019voxelmorph}, we take the output displacement field of \eac{DRN} as initialization for a gradient-descent based iterative optimization on each test scan individually, following the initial training process. 
This technique optimizes the same loss function that is used for \eac{DRN}. 
An overview of the complete method is given in \Cref{fig:network}.

\subsection{Evaluation}
\begin{figure}[tbp]
\centerfloat
\includegraphics[width=\textwidth]{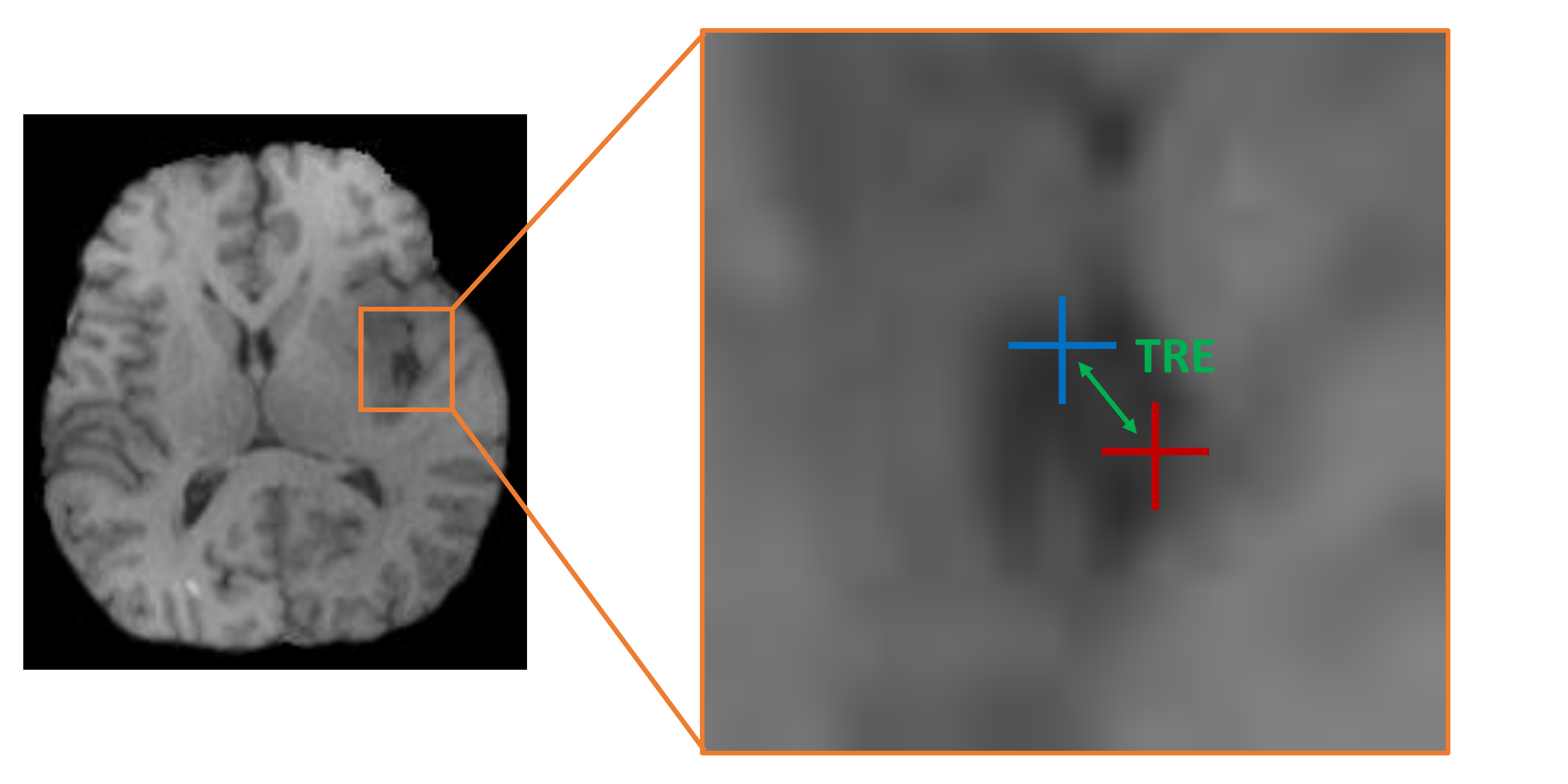}
\caption{Illustration of \eacf{TRE} evaluation between expert landmark annotation (red) and warped landmark after image registration (blue).
The green arrow indicates the Euclidean distance between the two points.
For simplicity, the illustration is in 2D, while the actual evaluation is performed in 3D space.}
\label{fig:landmarks}
\end{figure}

We evaluate all experiments by calculating the mean \eac{TRE} between the expert landmark annotations and the warped landmarks on the respective test sets.
We use the Euclidean distance to calculate the \eac{TRE}, as illustrated in \Cref{fig:landmarks}.

%% file: body/03_experiments_and_results.tex
\section{Experiments and Results}
We perform experiments using different loss functions on two brain \eac{MRI} datasets. 
Additionally, we implement two reference methods for comparison.

\subsection{Data and Pre-Processing} 
Training and evaluation are performed on two brain tumor datasets comprising 419 paired exams in total: 

\eac{PP}: Contains 300 pairs of brain MR exams with one timepoint before and one timepoint after tumor resection. 
It is comprised of the official training+validation dataset  of \eac{BraTS-Reg} \cite{baheti2021brain} with 160 cases and three datasets collected at \emph{Klinikum rechts der Isar (TUM)} with 49, 30, and 61 cases, respectively. 

\eac{PI}: Contains 119 pairs of brain MR exams with one timepoint before and one timepoint during tumor surgery and originates again from \emph{Klinikum rechts der Isar (TUM)}. 

Each exam comprises either three or four of the MR sequences \eac{T1}, \eac{T1CE}, \eac{T2}, \eac{FLAIR}. 
We preprocess the data using the \emph{BraTS toolkit} \cite{kofler2020brats}, which includes reorientation into the same coordinate system, rigid co-registration to a brain atlas as well as brain extraction. 
Additionally, intensity normalization is applied to all images.
For testing, we use 30 cases for \eac{PP} and 20 cases for \eac{PI}. For each test set, six landmarks annotated by clinical experts are provided for each case, accumulating to 180 landmarks on 30 patients for \eac{PP} and 120 landmarks on 20 patients for \eac{PI}.

\subsection{Similarity Metrics}
We demonstrate the simultaneous optimization strategy on the similarity metrics \eac{MSE} and \eac{NCC}, which are widely used for medical image registration tasks \cite{balakrishnan2019voxelmorph}. 

\subsection{Training and Testing} 
We perform multi-channel training and testing for \eac{DRN} and \eac{IO}. 
For \eac{PP}, this includes all four sequences, while for \eac{PI}, the sequences \eac{T1}, \eac{T1CE} and \eac{FLAIR} are available. 
To overcome this imbalance, we train separate networks for \eac{PP} and \eac{PI}.
We split the \eac{PP} dataset into training (80\%), validation (10\%), and test (10\%) and select the model for testing that shows a minimal loss on the validation set.
Since we have much fewer cases available for the experiments on the \eac{PI} dataset, we waive the validation set here and decide on a fixed number of 600 training epochs in all experiments.
At test time, the \eac{DRN}'s output deformation field serves as input to the 30 iterations of \eac{IO}.
For each dataset, we compare the combined loss of \eac{MSE}+\eac{NCC} against the individual addends, both for the initial \eac{DRN} training as well as the \eac{IO}.
Therefore, we exhaustively explore loss combinations for \eac{DRN} and \eac{IO}, as depicted in \Cref{tab:results_table}.

The implementation is inspired by \cite{balakrishnan2019voxelmorph} using \emph{Pytorch} 1.8.1 \cite{Paszke_PyTorch_An_Imperative_2019}.
We use learning rates $1e$-$4$ and $1e$-$3$ for \eac{DRN} and \eac{IO}, respectively. 
Image similarity metrics are weighted equally in the loss function, summing up to $1$ for the similarity term.
Smoothness regularization weight is set to $1.0$ for all experiments for a fair comparison.

\subsection{Reference Methods}
We compare our approach with two established methods: a deformable \emph{SyN} registration by \eac{ANTs} \cite{avants2009advanced}, and \eac{VM} \cite{balakrishnan2019voxelmorph}, both using the respective T1-weighted sequences.
We implement two variants of \eac{VM} \cite{balakrishnan2019voxelmorph}, available at \cite{voxelmorph2023github}, that employ \eac{MSE} and \eac{NCC} as respective  image similarity metrics. 
Therefore, we use default learning rate $1e$-$4$ and recommended smoothness regularization weights of $0.02$ and $1.5$ for \eac{MSE} and \eac{NCC}, respectively.
Likewise, \eac{ANTs} \emph{SyN} \cite{avants2009advanced} is implemented using default parameters. 
Since we opt for deformable registration without prior rigid/affine alignment, we apply the same for \eac{VM} and \eac{ANTs} \emph{SyN}.

\newpage
\subsection{Results}
\Cref{tab:results_table} shows quantitative results on both datasets.
The \eacp{TRE} indicate that for \eac{PP}, \eac{DRN} only with \eac{NCC} loss achieves the lowest mean error.
\input{body/03_results-table}
\input{body/03_stat-table}
When coupling \eac{DRN} with \eac{IO}, for all possible combinations, mean \eac{TRE} is always lowest when using \eac{MSE}+\eac{NCC} loss during \eac{IO}. 
On \eac{PI}, lowest mean \eac{TRE} for \eac{DRN} only is achieved when combining \eac{MSE} and \eac{NCC} in the loss function. 
When adding \eac{IO}, \eac{DRN} with \eac{NCC} loss coupled with \eac{IO} using \eac{MSE}+\eac{NCC} loss shows lowest mean \eac{TRE}.
Statistical comparisons based on \textit{Paired Samples T-Tests} are provided in \Cref{tab:t-test}.

Moreover, we have evaluated the experiments with respect to their \textit{hit rate} curves according to \cite{waldmannstetter2023framing}, see \Cref{fig:hit_miss}, showing the behavior of the different methods with respect to registration accuracy with increased tolerance.
\input{body/03_hitmiss-fig}

Evaluating performance with respect to running time, we determine that the runtime mainly depends on the implementation of the respective similarity metric. 
In our case, \eac{NCC} loss is significantly slower than \eac{MSE}. 
When combining the two metrics in the loss function, there is barely any difference in runtime compared to \eac{NCC} alone, since the runtime of \eac{MSE} is overall negligible.
This finding will probably change when the respective implementation differs and/or other similarity metrics are used.

\Cref{tab:results_overview} shows a comparison of the best \eac{DRN}+\eac{IO} with results achieved by reference methods \eac{VM} and \eac{ANTs}.
\input{body/03_overview-table}
For \eac{PP}, \eac{DRN}+\eac{IO} shows lowest mean \eacp{TRE}, while for \emph{PI}, this is the case for \eac{ANTs} \emph{SyN}.

Our findings are in line with how experts visually perceive the registration quality:
In a blinded evaluation of registration methods, three expert radiologists independently picked the \eac{MSE}+\eac{NCC} registration as their favorite.
Asked to provide reasoning for their choice, experts cited better registration quality around ventricles and fewer artifacts, such as unrealistically deformed tissue. \Cref{fig:qualitative} shows sample qualitative results on \eac{PI}, illustrating these findings.
\input{body/03_qualitative-fig}

%% file: body/03_results-table.tex
\begin{table}[tbp]
    \caption{
    Mean \eacp{TRE} in $mm$ with standard deviation using different losses for \eac{DRN} and \eac{DRN}+\eac{IO} on the Pre-Post-OP and Pre-Intra-OP test sets.
    Shown are the results on all possible combinations of loss functions.
    The best result in each category is highlighted. 
    Statistical comparisons are provided in \Cref{tab:t-test}. Hit rate curves \cite{waldmannstetter2023framing} are shown in \Cref{fig:hit_miss}.
    }
    \centering
    \scriptsize
    \begin{tblr}{colspec={lcr},colspec={Q[1.0,c] Q[1.0,c] Q[1.0,c] Q[1.0,c]}}
    \textbf{Loss [DRN]} & \textbf{Loss [IO]} & \textbf{Pre-Post-OP} & \textbf{Pre-Intra-OP} \\
    \hline
    MSE & (not applied)   & 2.40$\pm$1.56 & 3.48$\pm$3.45 \\
    MSE & MSE & 2.21$\pm$1.56 & 3.27$\pm$3.46 \\
    MSE & NCC & 1.81$\pm$1.45 & 3.24$\pm$3.39 \\
    MSE & MSE+NCC & \textbf{1.76$\pm$1.36} & \textbf{3.17$\pm$3.39} \\
    \hline
    NCC & (not applied)   & 2.18$\pm$1.52 & 3.31$\pm$2.68 \\
    NCC & MSE & 2.06$\pm$1.50 & 2.96$\pm$2.78 \\
    NCC & NCC & 1.80$\pm$1.44 & 2.71$\pm$2.79 \\
    NCC & MSE+NCC & \textbf{1.77$\pm$1.36} & \textbf{2.68$\pm$2.72} \\ 
    \hline
    MSE+NCC & (not applied) & 2.19$\pm$1.49 & 3.07$\pm$2.96 \\
    MSE+NCC & MSE & 2.07$\pm$1.48 & 2.94$\pm$3.05 \\
    MSE+NCC & NCC & 1.80$\pm$1.44 & 2.86$\pm$2.95 \\
    MSE+NCC & MSE+NCC & \cellcolor{blue!25}\textbf{1.73$\pm$1.34} & \textbf{2.76$\pm$2.89}
    \end{tblr}
    \label{tab:results_table}
\end{table}

%% file: body/03_stat-table.tex
\begin{table}[tbp]
\caption{\emph{P-values} and 95\% \emph{Confidence Intervals (CI)} of \emph{Paired Samples T-Tests} on the \emph{TRE} of competitive methods for the Pre-Post-OP and the Pre-Intra-OP datasets. 
Methods are given by declaring the used losses for \emph{DRN} and \emph{IO}, respectively.}
\centering
\scriptsize
\begin{tblr}{colspec={lcr},colspec={Q[1.0,c] Q[1.0,c] Q[1.0,c] Q[1.0,c]}}
\textbf{Method 1} & \textbf{Method 2} & \textbf{p-value [CI] (Pre-Post-OP)} & \textbf{p-value [CI] (Pre-Intra-OP)} \\
\hline
DRN (NCC) & DRN (MSE+NCC) & 0.85 [-0.09; 0.07] & 0.05 [0.00; 0.47] \\
DRN (MSE)\newline+ IO (NCC) & DRN (MSE)\newline+ IO (MSE+NCC) & 0.13 [-0.01; 0.10] & 0.23 [-0.04; 0.16] \\
DRN (NCC)\newline+ IO (NCC) & DRN (NCC)\newline+ IO (MSE+NCC) & 0.22 [-0.02; 0.08] & 0.50 [-0.06; 0.13] \\
DRN (MSE+NCC)\newline+ IO (NCC) & DRN (MSE+NCC)\newline+ IO (MSE+NCC) & \textbf{0.03} \textbf{[0.01; 0.13]} & \textbf{0.03} \textbf{[0.01; 0.19]} \\
\end{tblr}
\label{tab:t-test}
\end{table}

%% file: body/03_hitmiss-fig.tex
\begin{figure}[tbp]
         \centering
         \begin{subfigure}[b]{0.45\textwidth}
             {\includegraphics[width=\textwidth]{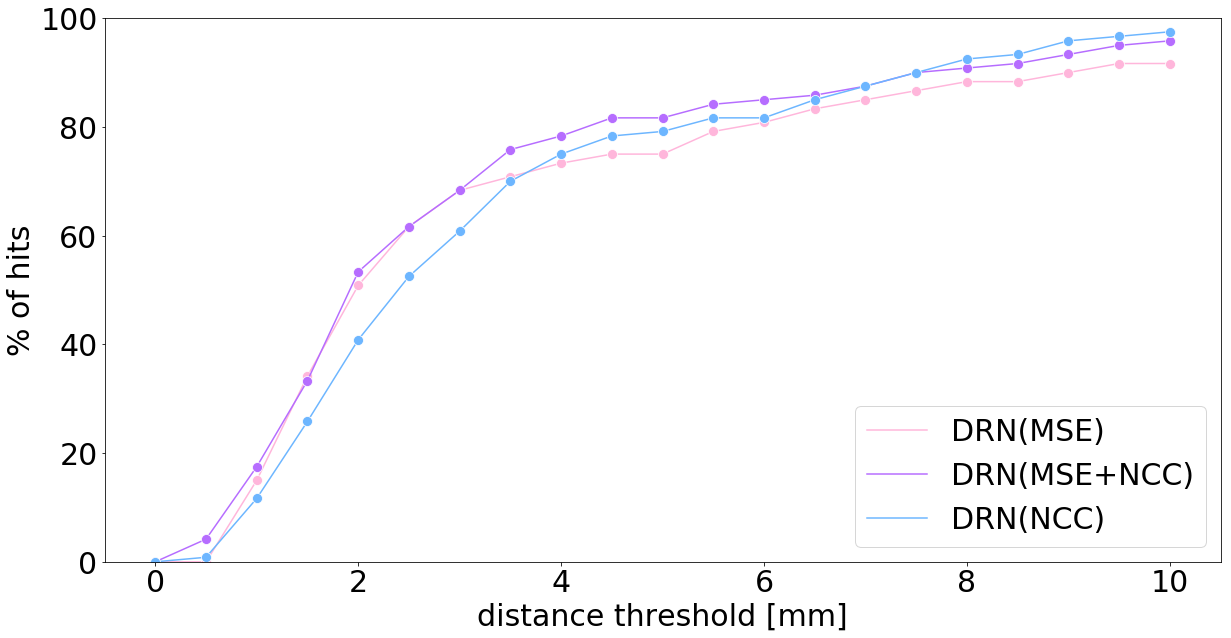}}
             \caption{PI: Percentage of \emph{hits} for three \emph{DRN} using different losses.}
             \label{fig:PI_groupDRN}
         \end{subfigure}
         \hspace{1mm}
         \begin{subfigure}[b]{0.45\textwidth}
             {\includegraphics[width=\textwidth]{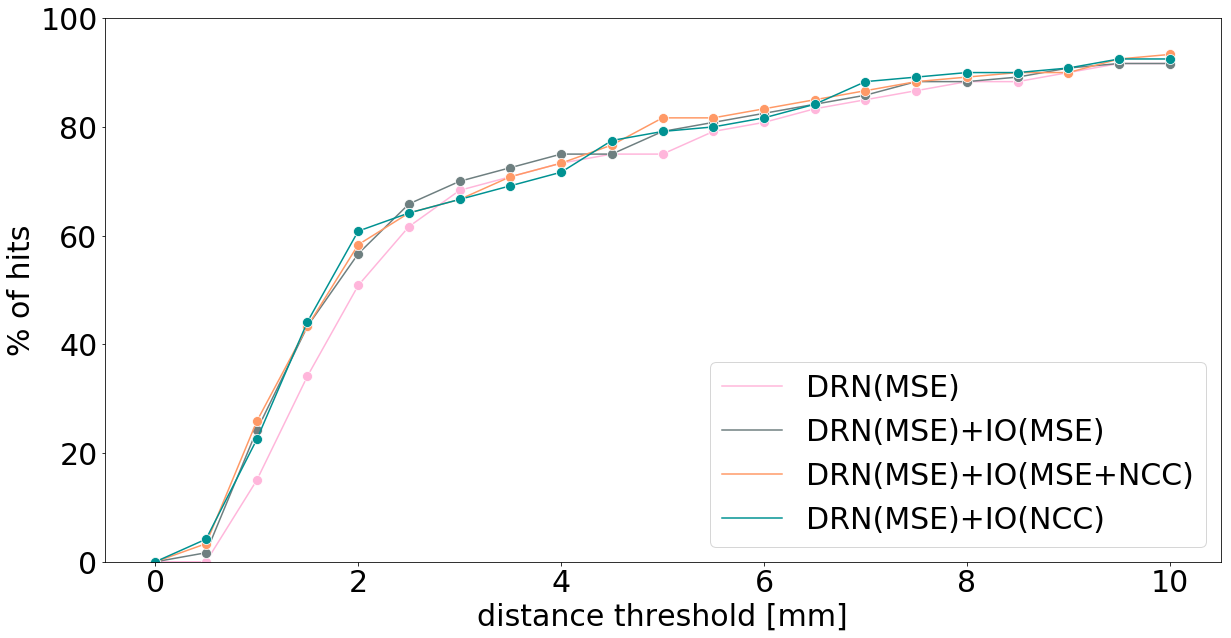}}
             \caption{PI: Percentage of \emph{hits} for \emph{DRN} using \emph{MSE} and different \emph{IO} losses.}
             \label{fig:PI_groupMSE}
         \end{subfigure}
         \hspace{1mm}
         \begin{subfigure}[b]{0.45\textwidth}
             {\includegraphics[width=\textwidth]{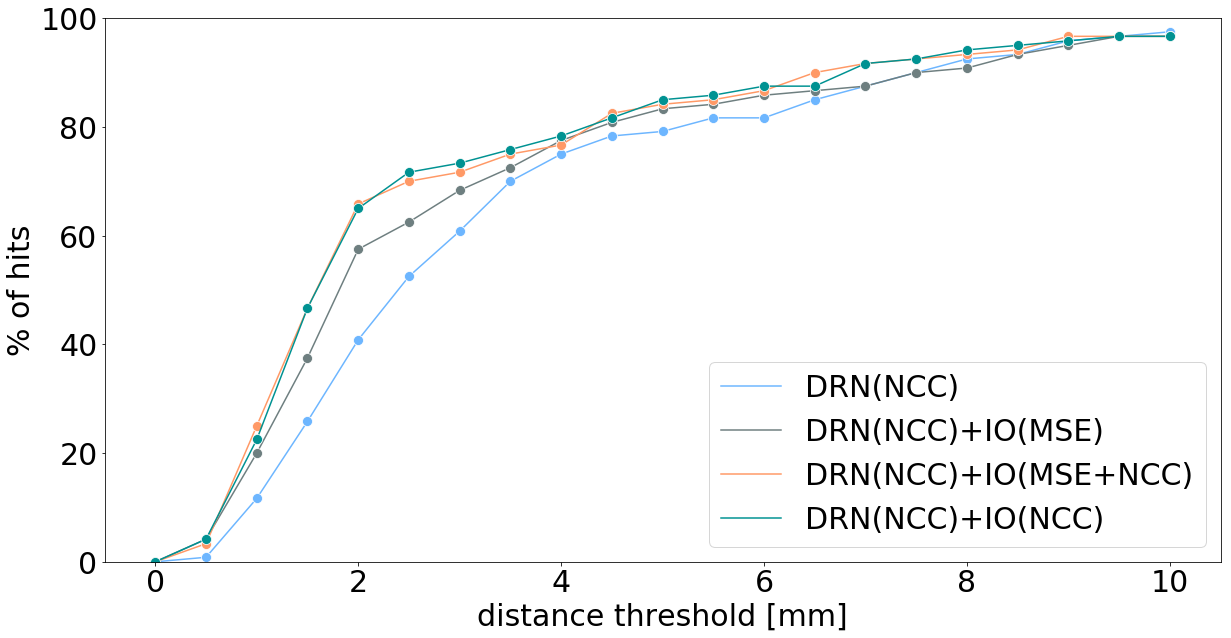}}
             \caption{PI: Percentage of \emph{hits} for \emph{DRN} using \emph{NCC} and different \emph{IO} losses.}
             \label{fig:PI_groupNCC}
         \end{subfigure}
         \hspace{1mm}
         \begin{subfigure}[b]{0.45\textwidth}
             {\includegraphics[width=\textwidth]{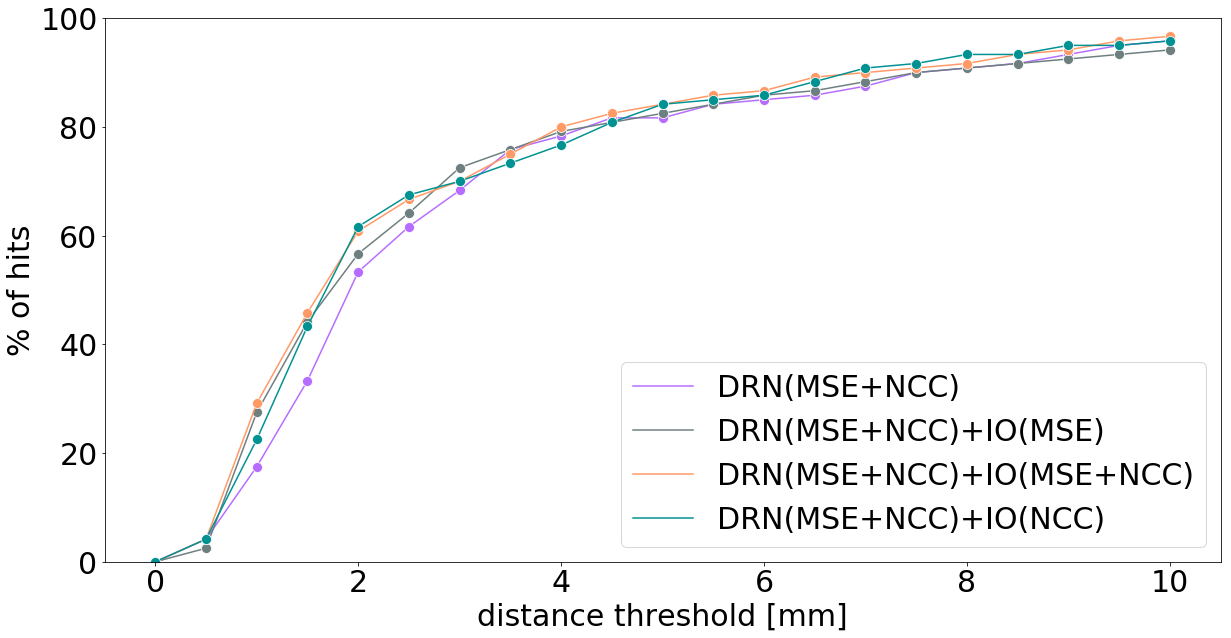}}
             \caption{PI: Percentage of \emph{hits} for \emph{DRN} using \emph{MSE+NCC} and different \emph{IO} losses.}
             \label{fig:PI_groupMSE+NCC}
         \end{subfigure}
         \hspace{1mm}
         \begin{subfigure}[b]{0.45\textwidth}
             {\includegraphics[width=\textwidth]{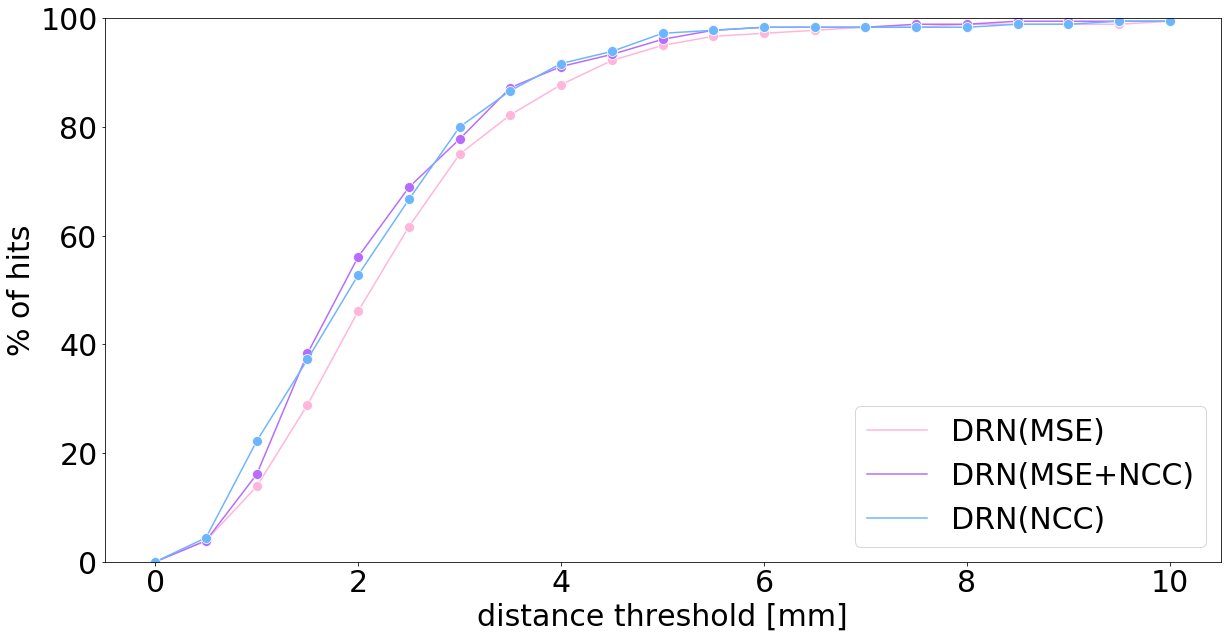}}
             \caption{PP: Percentage of \emph{hits} for three \emph{DRN} using different losses.}
             \label{fig:PP_groupDRN}
         \end{subfigure}
         \hspace{1mm}
         \begin{subfigure}[b]{0.45\textwidth}
             {\includegraphics[width=\textwidth]{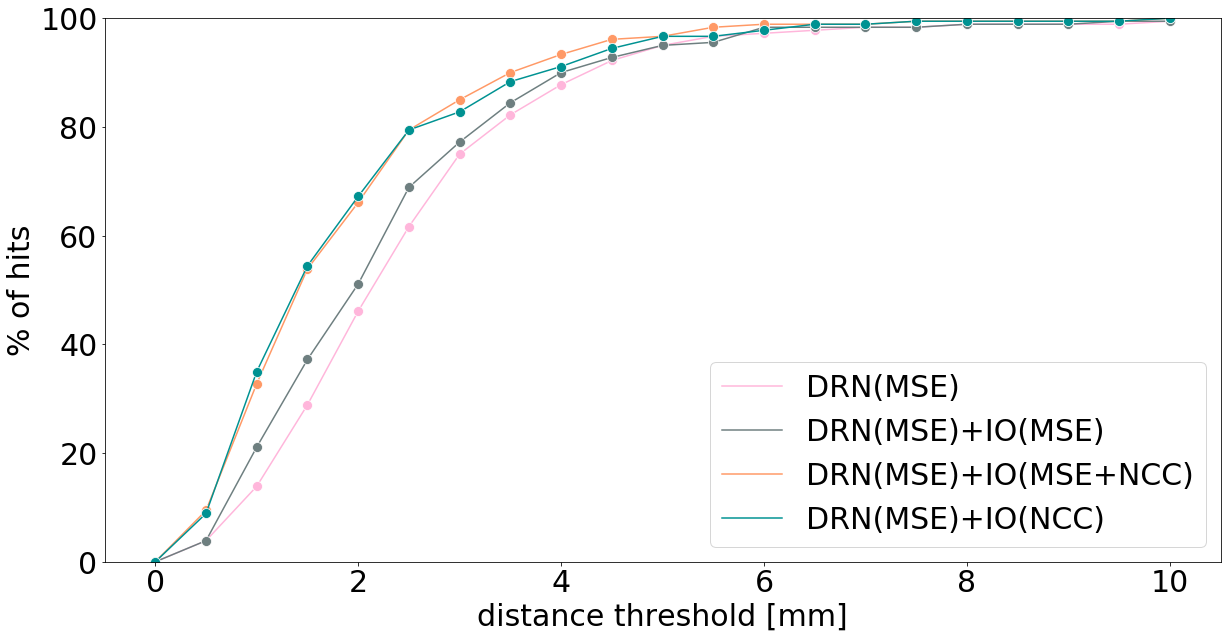}}
             \caption{PP: Percentage of \emph{hits} for \emph{DRN} using \emph{MSE} and different \emph{IO} losses.}
             \label{fig:PP_groupMSE}
         \end{subfigure}
         \hspace{1mm}
         \begin{subfigure}[b]{0.45\textwidth}
             {\includegraphics[width=\textwidth]{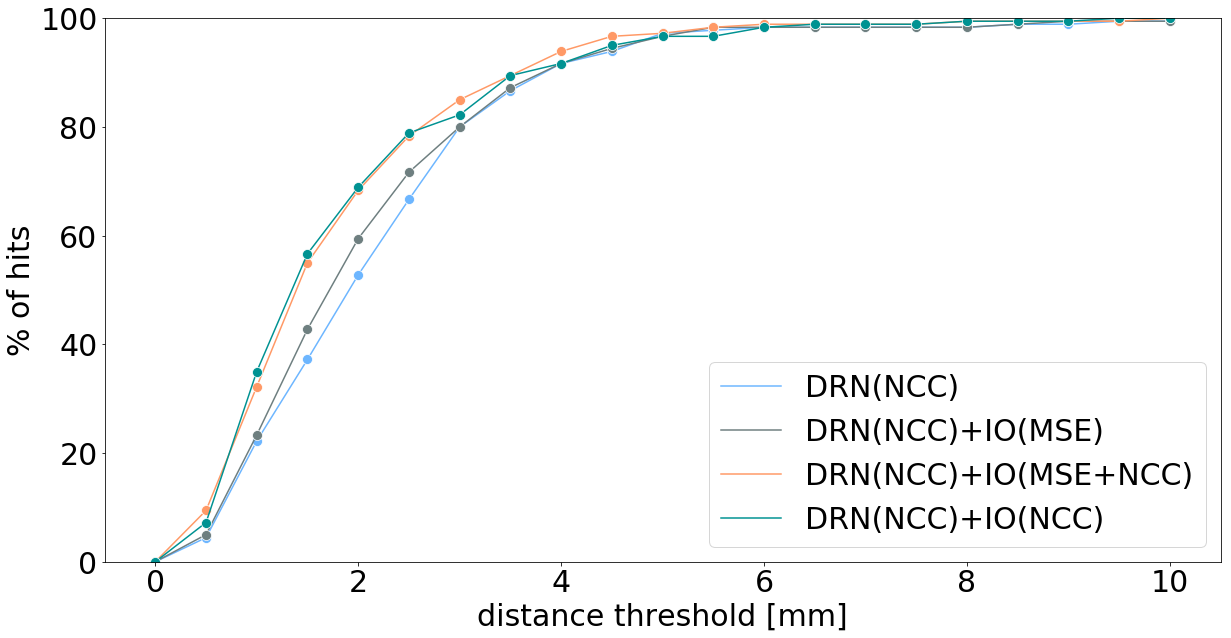}}
             \caption{PP: Percentage of \emph{hits} for \emph{DRN} using \emph{NCC} and different \emph{IO} losses.}
             \label{fig:PP_groupNCC}
         \end{subfigure}
         \hspace{1mm}
         \begin{subfigure}[b]{0.45\textwidth}
             {\includegraphics[width=\textwidth]{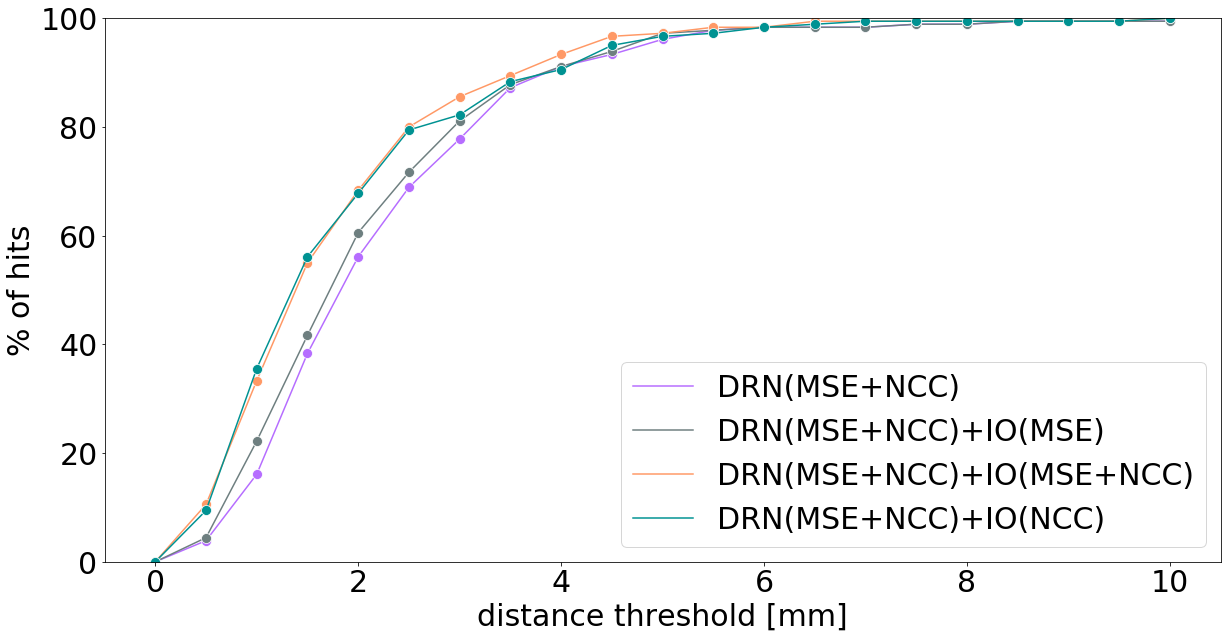}}
             \caption{PP: Percentage of \emph{hits} for \emph{DRN} using \emph{MSE+NCC} and different \emph{IO} losses.}
             \label{fig:PP_groupMSE+NCC}
         \end{subfigure}
         \caption{Hit rate curves \cite{waldmannstetter2023framing} on PI and PP, respectively. 
         \emph{Hit} denotes that a warped landmark lies within a certain distance threshold to the respective expert landmark annotation.
         Here, evaluation thresholds are set every $0.5 mm$, \emph{hit} percentages in between are interpolated.}
         \label{fig:hit_miss}
\end{figure}

%% file: body/03_overview-table.tex
\begin{table}[tbp]
    \caption{
    Comparison with reference methods.
    Mean \eac{TRE} in $mm$ with standard deviation on the Pre-Post-OP and Pre-Intra-OP datasets for the best \eac{DRN}+\eac{IO} compared to \eac{VM} and \acs{ANTs} \emph{SyN}.
    The proposed method is on par with established reference registration methods.
    }
    \centering
    \scriptsize
    \begin{tblr}{colspec={lcr},colspec={Q[1.0,c] Q[1.0,c] Q[1.0,c] Q[1.0,c]}}
    \textbf{Method} & \textbf{Pre-Post-OP} & \textbf{Pre-Intra-OP} \\
    \hline
    DRN+IO & 1.73$\pm$1.34 & 2.68$\pm$2.72  \\
    VM (MSE) & 2.98$\pm$2.11 & 3.66$\pm$2.90  \\
    VM (NCC) & 2.09$\pm$1.54 & 2.37$\pm$2.37  \\
    ANTs SyN & 2.02$\pm$1.49 & 1.81$\pm$1.17
    \end{tblr}
    \label{tab:results_overview}
\end{table}

%% file: body/03_qualitative-fig.tex
\begin{figure}[tbp]
\centering
\includegraphics[width=\textwidth]{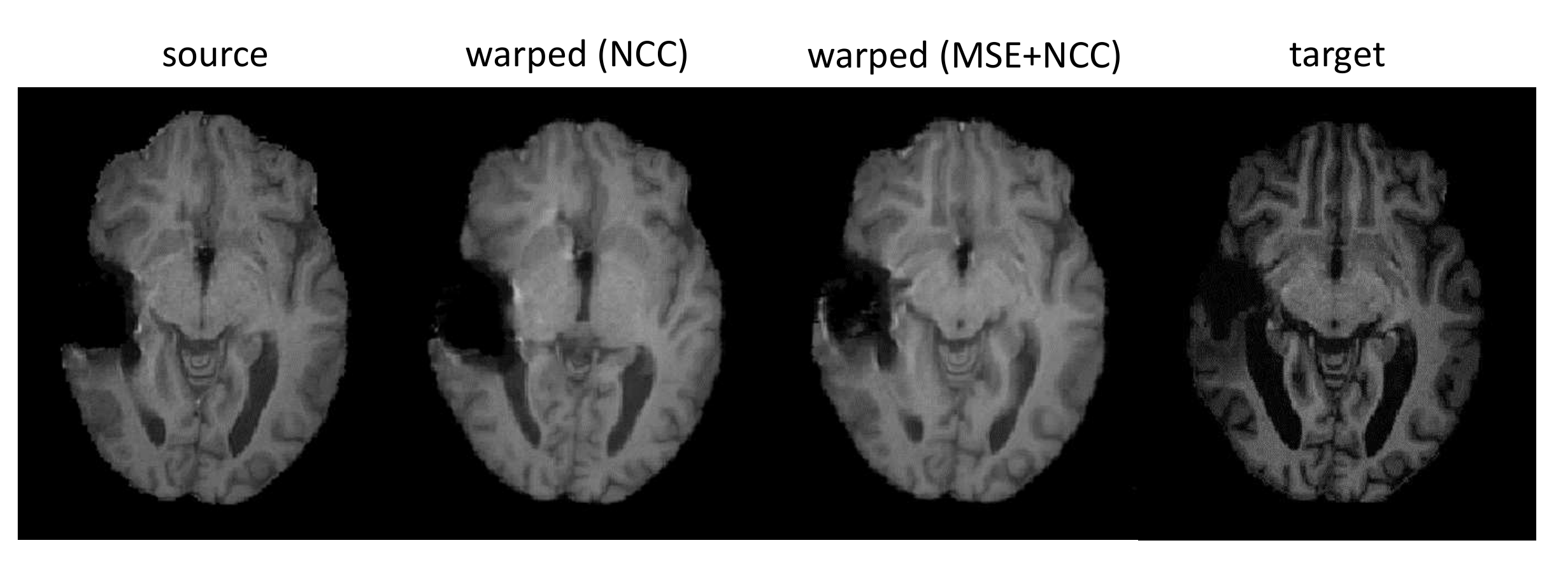}
\caption{Sample registration results on the Pre-Intra-OP dataset. Registration with combined losses \eac{MSE}+\eac{NCC} shows improved registration in comparison to registration with \eac{NCC} only, especially nearby the tumor and the resection cavity, respectively.}
\label{fig:qualitative}
\end{figure}

%% file: body/04_discussion.tex
\section{Discussion}
The contribution of this work is to simultaneously optimize multiple image similarity metrics for image registration tasks in a \eac{DL} setting.
For demonstration purposes, we combine the well-established metrics \eac{MSE} and \eac{NCC}.
We evaluate \eacp{DRN} with and without \eac{IO} in two challenging multi-modal 3D registration settings, namely pre- to post-operative and pre- to intra-operative glioma \eac{MRI}.

\Cref{tab:results_table} shows that combining \eac{MSE} and \eac{NCC} losses consistently improves \eac{IO} performance in both datasets.
When applied to \eac{DRN} only, the combination still performs best for \eac{PI} while showing comparable results to \eac{NCC} for \eac{PP}. 
Furthermore, \Cref{tab:results_overview} illustrates that our method achieves competitive results compared to established reference methods.
Moreover, for \eac{PP}, the proposed approach outperforms all reference methods. 
For \eac{PI}, \eac{ANTs} \emph{SyN} performs best.
A potential explanation might be lower registration quality introduced by the \eac{FLAIR} images. 
When using \eac{T1} images only for training and testing, the results are getting more comparable, see Table S1 in the supplementary material. 
Also, \eac{DRN} as well as \eac{VM} would likely benefit from a bigger training dataset.
When performing an additional experiment using the Elastix toolbox \cite{klein2009elastix} for single- vs multi-metric registration, we can observe that combining two metrics improves performance for \eac{PI}, see Table S2 in the supplementary material.

Combining \eac{MSE} and \eac{NCC} can help to improve the alignment of the source image to the target image. The qualitative assessment by clinical experts supports these findings.

There are several limitations to this study. 
The evaluation is performed on rather small test sets of 20 for \eac{PI} and 30 for \eac{PP}.
The used datasets are fairly specific, so for generalization purposes, an extension toward other registration tasks promises to yield additional insights.
Here, we focus on multi-sequence training on all available MR sequences using a 4D implementation of the \eac{NCC} loss.
Training on different sequences separately will possibly indicate increased understanding on the respective influences.
Performing cross validation for model selection might improve stability in the results. 
Even though the regularization could be fine-tuned, we use a fixed weight on the deformation field in our approach.

%% file: body/05_conclusion.tex
\newpage
\section{Conclusion and Future Work}
We propose an optimization strategy -- here implemented by means of a simple summation -- that benefits registration performance with regard to \eac{TRE}.
Moreover, we conduct extensive comparisons against established reference methods.
Future work should investigate the addition of further similarity metrics such as \emph{Mutual Information}.
Besides, introducing different weights for the individual parts of the loss function (instead of equally weighting) might further enhance registration performance.

%% file: body/acknowledgement.tex
\section*{Acknowledgements}
Supported by Deutsche Forschungsgemeinschaft (DFG) through TUM International Graduate School of Science and Engineering (IGSSE), GSC 81.
BM, BW and FK are supported through the SFB 824, subproject B12.
BM acknowledges support by the Helmut Horten Foundation. Finally, we acknowledge
Andreas Poschenrieder and Anna Valentina Lioba Eleonora Claire Javid Mamasani for eye-opening insights.

%% file: body/supplement.tex
\section*{Supplementary Material}

\setcounter{table}{0}
\renewcommand{\thetable}{S\arabic{table}}

\begin{table}[!htbp]
    \caption{
    Mean \emph{TRE} in $mm$ with standard deviation using different losses for \emph{DRN} and \emph{DRN}+\emph{IO} on the Pre-Intra-OP test set.
    For training and testing, only \emph{T1} images are used here.
    }
    \centering
    \scriptsize
    \begin{tblr}{colspec={lcr},colspec={Q[1.0,c] Q[1.0,c] Q[1.0,c] Q[1.0,c]}}
    \textbf{Loss [DRN]} & \textbf{Loss [IO]} & \textbf{Pre-Intra-OP} \\
    \hline
    NCC & (not applied)   & 2.54$\pm$2.38 \\
    NCC & NCC & 1.98$\pm$2.02 \\
    NCC & MSE+NCC & 1.98$\pm$2.10 \\ 
    \hline
    MSE+NCC & (not applied) & 2.63$\pm$2.75 \\
    MSE+NCC & NCC & 2.21$\pm$2.38 \\
    MSE+NCC & MSE+NCC & 2.18$\pm$2.33
    \end{tblr}
    \label{tab:results_t1}
\end{table}

\begin{table}[!htbp]
    \caption{Mean \emph{TRE} in $mm$ with standard deviation on the Pre-Post-OP and Pre-Intra-OP datasets for single- and multi-metric registration using the Elastix toolkit. \emph{MS} denotes mean squares similarity metric, while \textit{CC} denotes cross-correlation similarity metric. For \emph{multi-metric}, \emph{MS} and \emph{CC} are combined.
    }
    \centering
    \scriptsize
    \begin{tblr}{colspec={lcr},colspec={Q[1.0,c] Q[1.0,c] Q[1.0,c] Q[1.0,c]}}
    \textbf{Similarity Metric} & \textbf{Pre-Post-OP} & \textbf{Pre-Intra-OP} \\
    \hline
    MS & 11.45$\pm$13.61 & 7.91$\pm$10.18  \\
    CC & 2.83$\pm$2.29 & 2.84$\pm$2.42  \\
    multi-metric & 2.89$\pm$2.34 & 2.72$\pm$2.07
    \end{tblr}
    \label{tab:elastix}
\end{table}